\documentclass[%
 reprint,
 amsmath,
 amssymb,
 aps,
 ,showkeys
]{revtex4-2}

\usepackage{graphicx}
\usepackage{dcolumn}
\usepackage{bm}
\usepackage[colorlinks=true,citecolor=blue]{hyperref}
\usepackage{url}

\begin{document}
\preprint{}

\title{A study of topological quantities of lattice QCD by a modified DCGAN frame}

\author{Lin Gao}
\email{silvester\_gao@zju.edu.cn}
\affiliation{ School of Physics, Zhejiang University, Hangzhou, 310027, China}

\author{Heping Ying}
\email{hpying@zju.edu.cn}
\affiliation{ School of Physics, Zhejiang University, Hangzhou, 310027, China}

\author{Jianbo Zhang}
\email{jbzhang08@zju.edu.cn}
\affiliation{ School of Physics, Zhejiang University, Hangzhou, 310027, China}

\date{\today}

\begin{abstract}
A modified deep convolutional generative adversarial network (M-DCGAN) frame is proposed to study the N-dimensional (ND) topological quantities in lattice QCD based on the Monte Carlo (MC) simulations. We construct a new scaling structure including fully connected layers to support the generation of high-quality high-dimensional images for the M-DCGAN. Our results show that the M-DCGAN scheme of the Machine learning should be helpful for us to calculate efficiently the 1D distribution of topological charge and the 4D topological charge density compared with the case by the MC simulation alone.
\end{abstract}

\maketitle


\section{Introduction}
Machine learning (ML) is changing our world in many domains including physical statistics\cite{lecun2015deep}\cite{Carrasquilla2017}\cite{GoodfellowNIPS2014}\cite{goodfellow2020generative}. For example, the well-known ChatGPT has demonstrated powerful capabilities in chatting and has huge potential to improve the scientific research efficiency of physicists, such as helping physicists write code to draw histograms of physical statistics\cite{openai2023gpt4}\cite{chatgptreasoning}. ML can also be applied in lattice QCD. Simulating data for topological quantities of lattice QCD by traditional Monte Carlo (MC) methods is very time-consuming\cite{Wilson1974PRD}\cite{Gattringer2010}, so we try to use ML to speed up data simulation. Regarding the application of ML in Lattice QCD, there have been many studies. Sebastian J. Wetzel and Manuel Scherzer have used the ML to explore the phase transition locations of SU(2) lattice gauge theory \cite{wetzel2017machine} . Jan M Pawlowski and Julian M Urban have applied the generative adversarial networks to reduce autocorrelation times for the lattice simulations\cite{Pawlowski_2020}. Simone Bacchio et al. used ML based on Lüscher's perturbative results to learn trivializing maps in a 2D Yang-Mills theory\cite{bacchio2023learning} . Matteo Favoni and Andreas Ipp et al. propose Lattice gauge equivariant Convolutional Neural Networks to study lattice gauge theory\cite{Favoni2022}. These works demonstrate the broad application of the ML in lattice QCD, including phase transition studies in critical regions, correlation time reduction, field transformations and the approximation of gauge covariant function, etc. Our research focuses on constructing a ML framework to study topological quantities of lattice QCD and is committed to generating corresponding data based on a necessary amount of MC data to save computing times.

We use the generative machine learning model framework in this paper. The so-called generative model means that it does not just label the data, but learns the original training data and generates new data. Generative adversarial network (GAN) is a kind of generative model and it uses two components generator and discriminator which not only receive data, but also compete with each other internally to select better components, and it is different from using a single model to passively receive and digest data\cite{GoodfellowNIPS2014}\cite{goodfellow2020generative}. Next, we discuss the distribution generation of GAN. Assume that we now have some original data and the data obeys a certain Gaussian distribution. At this time, we can first write the Gaussian distribution probability density function with the mean and variance of the undetermined parameters, and then calculate the parameters based on the data. This probability density function can then be combined with a certain algorithm to obtain target Gaussian distribution that is the same as the distribution of the original data from the uniform distribution. If we package the probability density function and this algorithm as a function, we can start from the uniform distribution and get the Gaussian distribution of the original data through this function. As we know, in many other cases, the distribution of a certain original data has thousands or even millions of parameters, it will be very difficult for us to follow the above steps to get this new distribution from the uniform distribution. Therefore, it is necessary to use ML to find a function G(x) that can generate the distribution $\mathbb{P}_{g}$ which is the same or very similar to the distribution $\mathbb{P}_{r}$ of the real data from a known simple distribution $\mathbb{P}_{k}$. Different ML models have different methods to find this function G(x), and the GAN obtain this function G(x) through the following two quantities\cite{GoodfellowNIPS2014}\cite{goodfellow2020generative}
\begin{equation}
\mathrm{max}\left\{\mathbb{E}_{x\sim\mathbb{P}_{r}}[ln D(x)]+\mathbb{E}_{y\sim\mathbb{P}_{g}}[ln\left(1-D(y)\right)]\right\},
\end{equation}
and
\begin{equation}
\mathrm{max}\mathbb{E}_{z\sim\mathbb{P}_{k}}[ln(D(G(z)))],
\end{equation}
where D(x) is the discriminator function, $\mathbb{E}$ refers to the expectation and $\mathrm{max}\mathbb{E}$ is maximizing expectation. GAN uses these two formulas to find function G(x) from the known original data, so that the new data generated by G(x) has the same or very similar distribution as the original data. The topological charge density and topological charge distribution that we are going to discuss can all be regarded as distributions, so GAN can be used to generate related data. After GAN was proposed, there have been many variants, one of which is deep convolutional generative adversarial network (DCGAN)\cite{lecun1995convolutional}\cite{radford2015unsupervised}. The DCGAN introduces strided convolutions in discriminator and fractional-strided convolutions in generator. And it also uses batchnorm, ReLU activation and LeakyReLU activation. As a result, the DCGAN greatly improves the stability of GAN training as well as the quality of results.

\section{Data preparation}

\subsection{The lattice QCD}

QCD is a standard dynamical theory used to describe fermion quarks and gauge boson gluons in the strong interaction. The perturbation theory is invalid in the low energy regime of QCD\cite{GeorgiPolitzer1976} and a reliable non-perturbative method is desired. Therefore, it is necessary for us to find ways to construct a non-perturbative theory to deal with problems in QCD. Lattice QCD is a non-perturbative theory that can be used to study QCD based on first principles\cite{Wilson1974PRD}. In lattice QCD we use Euclidean spacetime instead of Minkowski spacetime, which helps the computer deal with real numbers instead of complex numbers. Therefore, the formulas we use default to the form in Euclidean spacetime. In addition, we need to discretize the physical quantities in continuous spacetime, and the discrete gauge action used is the Wilson gauge action\cite{Gattringer2010} 
\begin{equation}
{S_G}\left[ U \right] = \frac{\beta }{3}\mathop \sum \limits_{n \in {\rm{\Lambda }}} \mathop \sum \limits_{\mu  < \nu} {\mathop{\rm Re}\nolimits} {\rm{Tr}}\left[ {1 - {U_{\mu \nu}}\left( n \right)} \right],
\end{equation}
where $U_{\mu\nu}\left(n\right)$ is the plaquette and $\beta$ is the inverse coupling. Next, we need to introduce the topological quantities which are the topological charge density, the topological charge and the topological susceptibility. The topological charge density is written as
\begin{equation}
q\left(n\right)=\frac{1}{32\pi^2}\varepsilon_{\mu\nu\rho\sigma}Re{Tr{\left[F_{\mu\nu}^{clov}\left(n\right)F_{\rho\sigma}^{clov}\left(n\right)\right]}},
\end{equation}
where $F_{\mu\nu}^{clov}\left(n\right)$ is the clover improved lattice discretization of the field strength tensor $ F_{\mu\nu}\left(x\right)$ .The $F_{\mu\nu}^{clov}\left(n\right)$  can be described as
\begin{footnotesize}
\begin{equation}
\begin{array}{*{20}{c}}
F_{\mu\nu}^{clov}(n)=-\frac{i}{8a^2}\left[\left(C_{\mu\nu}\left(n\right)-C_{\mu\nu}^\dag(n)\right)-\frac{Tr\left(C_{\mu\nu}\left(n\right)-C_{\mu\nu}^\dag(n)\right)}{3}\right],
\end{array}
\end{equation}
\end{footnotesize}
where the clover can be expressed as
\begin{equation}
C_{\mu\nu}(n)=U_{\mu,\nu}(n)+U_{\nu,-\mu}(n)+U_{-\mu,-\nu}(n)+U_{-\nu,\mu}(n).
\end{equation}
The topological charge density appears in the anomalous axial vector current relation \cite{tHooft1986}, and provides a resolution of the ${U\left(1\right)}_A$ problem in QCD \cite{Gattringer2010}\cite{BelavinPolyakov1975}. In the path integral formalism of QCD, this ${U\left(1\right)}_A$ anomaly comes from the noninvariance of the quark field measure under the ${U\left(1\right)}_A$ transformation of quark fields. Furthermore, the topological charge is given as
\begin{equation}
Q_{top}=a^4\sum_{n\in\mathrm{\Lambda}} q\left(n\right),
\end{equation}
which is an integer in the continuous case \cite{atiyah1971index}. The topological charge density and the topological charge are important in understanding the property of the QCD vacuum\cite{tHooftPhysRevD1976}\cite{hooft2000monopoles}. Moreover, the topological susceptibility $\chi_t$ is expressed as
\begin{equation}
\chi_t\ =\frac{1}{V}\left\langle{Q_{top}}^2\right\rangle,
\end{equation}
where $V$ is the 4D volume. The Witten-Veneziano relation describes that the mass squared of $\eta^\prime$ meson is proportional to the topological susceptibility of pure gauge theory for massless quarks\cite{witten1979current}\cite{veneziano1979u}. In addition, the Wilson flow is introduced to improve the configuration\cite{Luscher2010}\cite{Zhang2010rn}\cite{Zou2018}.

\begin{figure}[htb]
\includegraphics[width=0.4\textwidth]{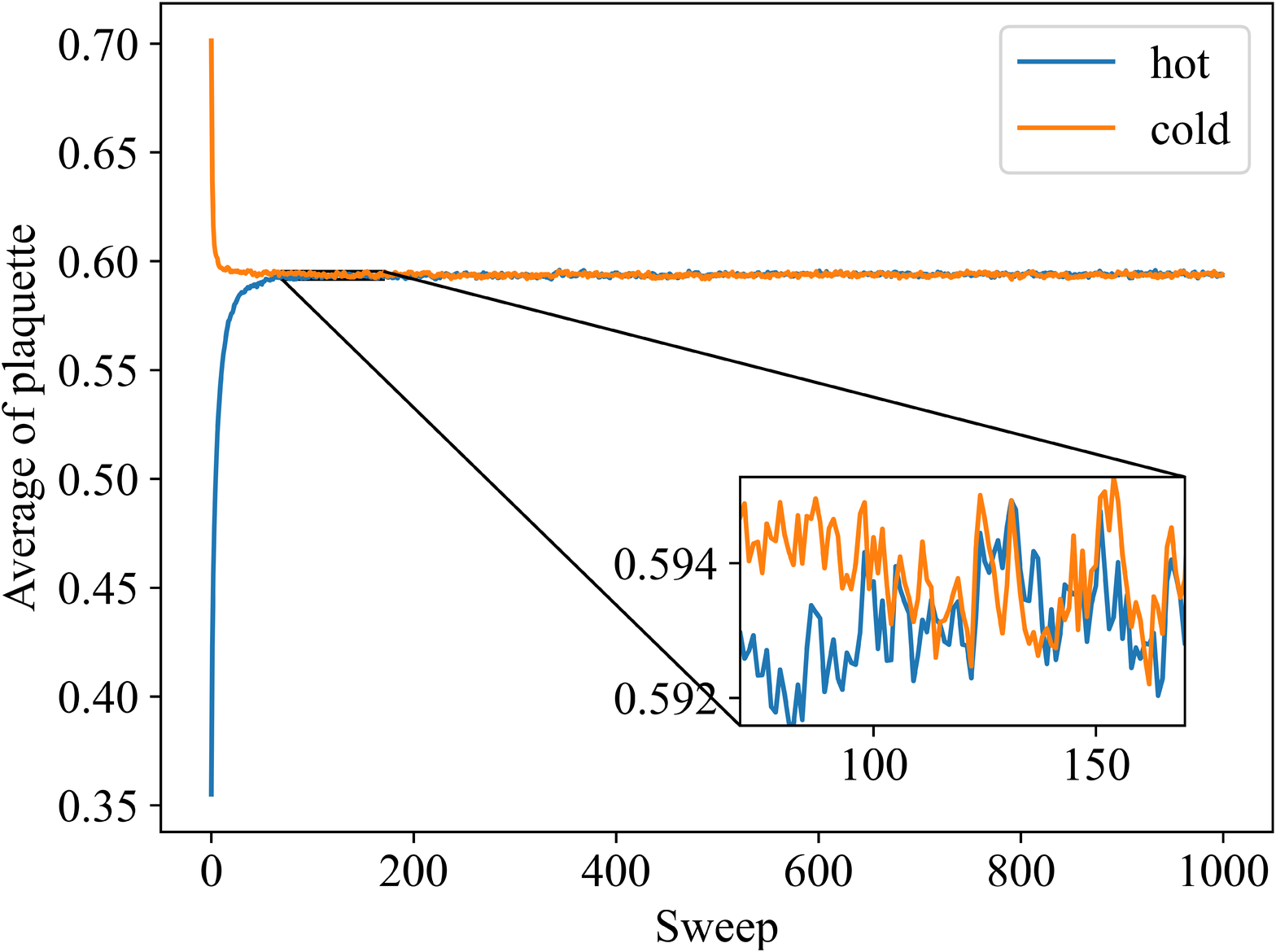}
\caption{\label{figure_1_ref}The evolution of the average of plaquette under different initial conditions.}
\end{figure}

The next part is the preparation of original data. We use the pseudo heat bath algorithm with the periodic boundary conditions to simulate configurations of lattice QCD and apply Wilson flow to smear configurations\cite{Gattringer2010}. Then we calculate topological charge density, topological charge and topological susceptibility from the configurations. The configurations are generated by the software Chroma on individual workstation\cite{Edwards_2005}. In detail, the updating steps are repeated 10 times for the visited link variable because the computation of sum of staples is costly and hot start have been applied. The Wilson flow step time $\varepsilon_f=0.01$ and total number of steps $N_{flow}=600$ are chosen. Moreover, we use the mean value of plaquette $\left\langle {\frac{1}{3}ReTr{U_{plaq}}} \right\rangle $ with $\beta=6.0$ to determine whether the system has reached equilibrium. It is found from Fig.~\ref{figure_1_ref} that the system has reached equilibrium after about 200 sweeps because the mean value of plaquette has evolved to a similar value starting from different initial conditions including cold start and hot start.
Furthermore, we need to calculate the integrated autocorrelation time which should be small. For the Markov sequence generated by MC, $X_i$ is a random variable and we could introduce the autocorrelation function
\begin{equation}
{C_X}\left( {{X_i},{X_{i + t}}} \right) = \left\langle {{X_i}{X_{i + t}}} \right\rangle  - \left\langle {{X_i}} \right\rangle \left\langle {{X_{i + t}}} \right\rangle .
\end{equation}
We note $C_X\left(t\right)\mathrm{=}C_X\left(X_i,\mathrm{\ } X_{i+t}\right)$ at equilibrium and the normalized autocorrelation function $\Gamma_X\left(t\right)=\frac{C_X\left(t\right)}{C_X\left(0\right)}$. We define the integral autocorrelation time as
\begin{equation}
{\tau _{X,int}}= \frac{1}{2} + \sum\limits_{t = 1}^N {{\Gamma _X}(t)} .
\end{equation}
We apply the topological charge in this article to calculate the integral autocorrelation time whose value is 0.416 which indicates that each of data can be regarded as independent because of $N/2\tau_{X,\mathrm{\ int\ }}>N$ \cite{Gattringer2010}, where a total of 1000 configurations are sampled with intervals of 200 sweeps.

In addition, the static QCD potential is used to set the scale and can be parameterized by
$V\left(r\right)=A+\frac{B}{r}+\sigma r$\cite{Gattringer2010}. The Sommer parameter $r_\mathrm{0}$ is defined as
\begin{equation}
\left(r^2\frac{dV\left(r\right)}{dr}\right)_{r=r_0}=1.65,
\end{equation}
and the Sommer parameter $\ r_\mathrm{0}\mathrm{\ =\ }0.49fm$ is used\cite{sommer2014scale} \cite{SOMMER1994}. The scales are shown in the Tab.~\ref{static_potential_scale}.

\begin{table}[htb]
\caption{\label{static_potential_scale}Setting scales through the static QCD potential.}
\begin{ruledtabular}
\begin{tabular}{cccccc}
\textrm{$lattice$}&
\textrm{$\beta$}&
\textrm{$a[fm]$}&
\textrm{$N_{cnfg}$}&
\textrm{${r_0}/a$}&
\textrm{$L[fm]$}\\
\colrule
$24\times{12}^3$ & 6.0&0.093(3)&1000&5.30(15)&1.11(3) \\
\end{tabular}
\end{ruledtabular}
\end{table}

\subsection{Machine learning scheme}
\begin{figure}[htb]
\includegraphics[width=0.4\textwidth]{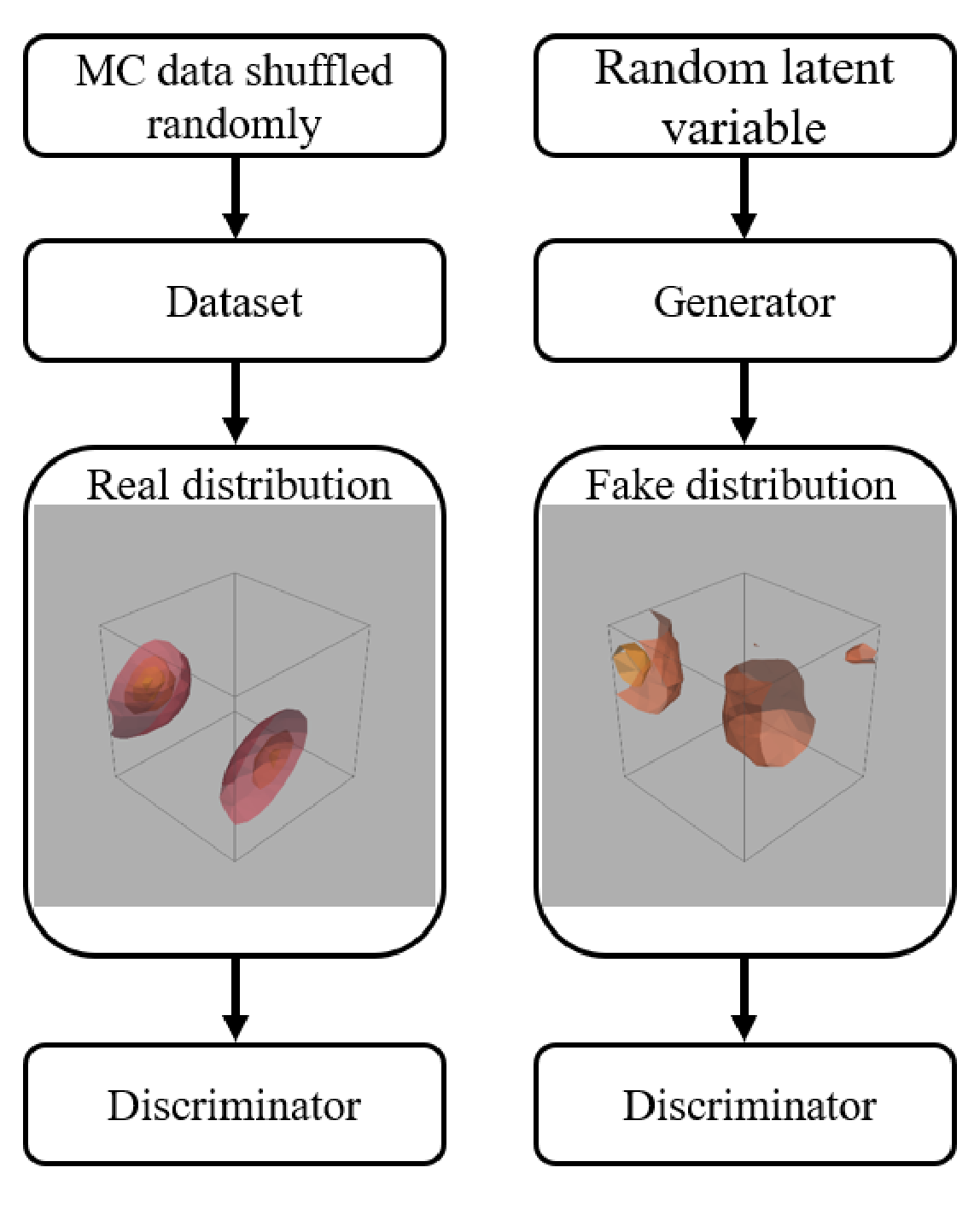}
\caption{\label{figure_2_ref}The overall structure of M-DCGAN.}
\end{figure}

For our purpose, the suitable ML models should be built to help study the characteristics of topological charge and topological charge density. For the large-sized tensor data generated in this paper, the DCGAN will not perform well due to too many times of transposed convolution. Therefore, a new frame of DCGAN is proposed to generate topological charge and topological charge density for our study. Compared with the original DCGAN frame, we construct a new scaling structure. Some fully connected layers have been added basically to our solution, so that the layers of transposed convolution can be reduced while processing large-size tensor. The overall structure of M-DCGAN is shown in Fig.~\ref{figure_2_ref}.

The structure of the generator for N-dimensional M-DCGAN (ND M-DCGAN) is explained in Tab.~\ref{generator_structure_tab} using 1D M-DCGAN as an example. For the ND model, the content of each layer needs to be adjusted according to the dimension. Moreover, the functions of various types of layers are as follows. The fully connected layers and transposed convolution are applied to amplify and reshape the input latent variable with normal distribution. The convolution is applied to scale down and reshape the previously enlarged data. This new scaling structure is beneficial for large tensor data. The batch normalization is exerted to improve generation. All layers use LeakyReLU activation except the output layer uses Tanh.

The input of generator for 1D M-DCGAN is a random latent variable tensor with shape [-1,100], and the output is a tensor with shape [-1,1,100], where -1 is an undetermined parameter. For example, we need to generate 800 topological charge values, the input shape is [8,100], and the output shape is [8,1,100]. Finally, the output needs to be reshaped into an 800-dimensional vector. These 800 data can form a distribution.

\begin{table}[htb]
\caption{\label{generator_structure_tab}The structure of the generator for 1D M-DCGAN. Layers 1-21 constitute the new scaling structure.}
\begin{ruledtabular}
\begin{tabular}{llr}
\textrm{Layer (type) }&
\textrm{Output Shape }&
\textrm{Parameter number }\\
\colrule
         Linear-1      &         $ [-1, 10000] $     & 1,000,000\\
       BatchNorm1d-2   &       $      [-1, 10000]$&          20,000\\
         LeakyReLU-3    &         $   [-1, 10000]$ &              0\\
            Linear-4          &        $[-1, 100] $      &1,000,000\\
       BatchNorm1d-5   &          $     [-1, 100]$&             200\\
         LeakyReLU-6    &            $  [-1, 100] $ &             0\\
            Linear-7          &          $[-1, 4]  $       &    400\\
       BatchNorm1d-8   &            $     [-1, 4] $ &             8\\
         LeakyReLU-9    &              $  [-1, 4]$    &           0\\
           Linear-10         &         $[-1, 100]   $   &       400\\
      BatchNorm1d-11  &           $     [-1, 100]$&             200\\
        LeakyReLU-12   &             $  [-1, 100]$ &              0\\
           Linear-13         &         $  [-1, 4]   $     &     400\\
      BatchNorm1d-14  &            $      [-1, 4] $&              8\\
        LeakyReLU-15   &              $   [-1, 4]   $   &         0\\
           Linear-16         &         $[-1, 100]  $        &   400\\
      BatchNorm1d-17  &           $     [-1, 100]$   &          200\\
        LeakyReLU-18   &             $  [-1, 100] $   &           0\\
           Linear-19        &         $[-1, 6400]  $       &640,000\\
      BatchNorm1d-20  &          $     [-1, 6400]$  &        12,800\\
        LeakyReLU-21   &            $  [-1, 6400]$   &            0\\
  ConvTranspose1d-22&             $ [-1, 128, 25]$&          98,304\\
      BatchNorm1d-23  &            $[-1, 128, 25]  $ &          256\\
        LeakyReLU-24   &          $ [-1, 128, 25] $   &           0\\
  ConvTranspose1d-25    &          $ [-1, 64, 50] $&         32,768\\
      BatchNorm1d-26       &       $ [-1, 64, 50] $   &         128\\
        LeakyReLU-27        &     $  [-1, 64, 50]  $   &          0\\
  ConvTranspose1d-28    &         $  [-1, 1, 100]$ &            256\\
             Tanh-29           &  $  [-1, 1, 100]    $       &    0\\

\end{tabular}
\end{ruledtabular}
\end{table}

The structure of the discriminator for ND M-DCGAN is described in Tab.~\ref{discriminator_structure_tab}, taking 1D M-DCGAN as an example. The convolution is used to scale down and reshape images. All layers use LeakyReLU except for the output layer. It is worth noting that sigmoid layer is placed in loss function BCEWithLogitsLoss in this article. Dropout is used in the discriminator due to the small number of training samples for topological charge density. The parameters in Tab.~\ref{discriminator_structure_tab} are explained as follows. The input shape for Conv1d-1 in Tab.~\ref{discriminator_structure_tab} is [-1,1,100], the shape of output is [-1,64,50] and the kernel size is 4, therefore the parameter number is $1 \times 64 \times 4 = 256$.

\begin{table}[htb]
\caption{\label{discriminator_structure_tab}The structure of the discriminator for 1D M-DCGAN.}
\begin{ruledtabular}
\begin{tabular}{llr}
\textrm{Layer (type) }&
\textrm{Output Shape }&
\textrm{Parameter number }\\
\colrule
           Conv1d-1             &$  [-1, 64, 50]        $ &    256\\
         LeakyReLU-2       &    $    [-1, 64, 50]      $     &    0\\
           Dropout-3              & $[-1, 64, 50]        $    &   0\\
            Conv1d-4        &     $ [-1, 128, 25]     $   &  32,768\\
         LeakyReLU-5        &   $   [-1, 128, 25]$  &             0\\
           Dropout-6         &    $ [-1, 128, 25]$   &            0\\
            Linear-7              &   $   [-1, 1]    $  &     3,200\\
\end{tabular}
\end{ruledtabular}
\end{table}

The optimizer is the Adam, which comprehensively deals with the variable learning rate and momentum proposed by Kingma and Ba\cite{kingma2014adam}. We choose the Adam as a suitable optimizer because it requires less memory, automatically adjusts the learning rate and limits the update step size to a general range. As well as it is very suitable for large-scale data and parameter scenarios.

As a result, the M-DCGAN models can generate 1D topological charge and 4D topological charge density directly to save computational times after training. The 1D M-DCGAN and 4D M-DCGAN use the same structure described above except for different dimensions and realize unsupervised generation without labels. We adopt the mini-batch method in the training phase. In addition, some programs are based on Pytorch\cite{PASZKE2019PyTorch}.

\section{Numerical results}
\subsection{The distribution of the topological charge}
\begin{figure}[htb]
\includegraphics[width=0.5\textwidth]{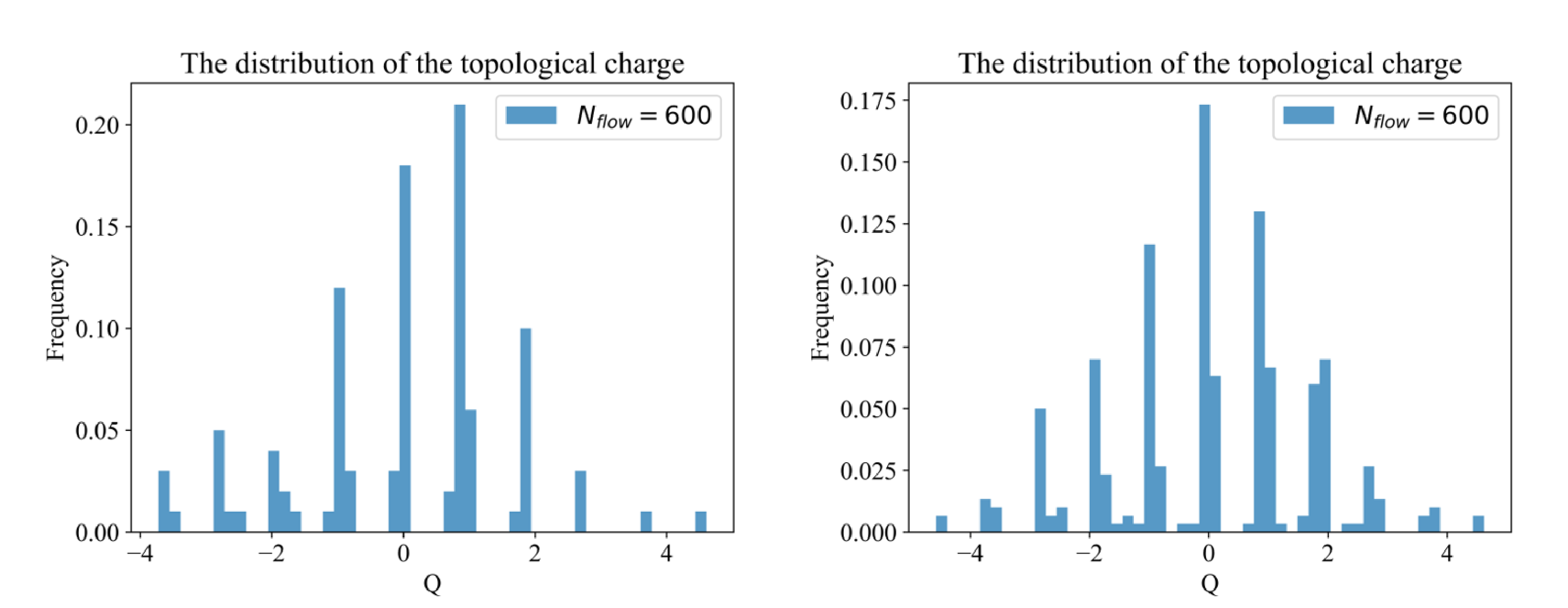}
\caption{\label{figure_3_ref}The distribution of the topological charge based on the MC with Wilson flow time steps $N_{flow}=600$. The numbers of topological charge are 100 for the left sub-picture and 300 for the right sub-picture.}
\end{figure}

First, we present the distribution of the topological charge simulated by the MC with Wilson flow for the cases where the number of the topological charge is 100 and 300 respectively. We obtain from the right sub-picture of Fig.~\ref{figure_3_ref} that the fourth root of topological susceptibility is ${\chi_t}^{1/4}=191.8\pm3.9MeV$ in terms of $N_{flow}=600$ and a reference calculation in the literature is ${\chi_t}^{1/4}=191\pm5MeV$\cite{DELDEBBIO2005topological}. We use the jack-knife to analyze the error of data\cite{LiuChuan2017}. Moreover, we can find that the topological charge Q is distributed near integers from Fig.~\ref{figure_3_ref} being consistent with the aforementioned conclusion that Q is an integer in the continuum. Furthermore, the topological charge distribution should be symmetric about the origin. However, we can see that the distribution in the left panel of Fig.~\ref{figure_3_ref} does not have good symmetry due to the poor statistics of data. Therefore, it is better to improve the distribution of data by increasing their statistics. In MC simulations, increasing the amount of data will result in a rapid increase in storage usage and time cost. Luckily, the ML model can almost avoid this problem. The ML model can immediately generate a corresponding data to improve the accuracy of the results once it has finished training. Next, we will discuss the details of two methods, the MC with Wilson flow and ML with the M-DCGAN scheme, as well as apply a mixture of two methods to generate the distribution of topological charge more efficiently.

For MC, we simulated 1600 configurations of lattice gauge field and calculated the topological charge distribution and the fourth root of the topological susceptibility ${\chi_t}^{1/4}=190.9\pm1.7MeV$ from these configurations with 25 CPU cores in our computations, the details of which are shown in Tab.~\ref{MC and ML}.

\begin{figure}[htb]
\includegraphics[width=0.5\textwidth]{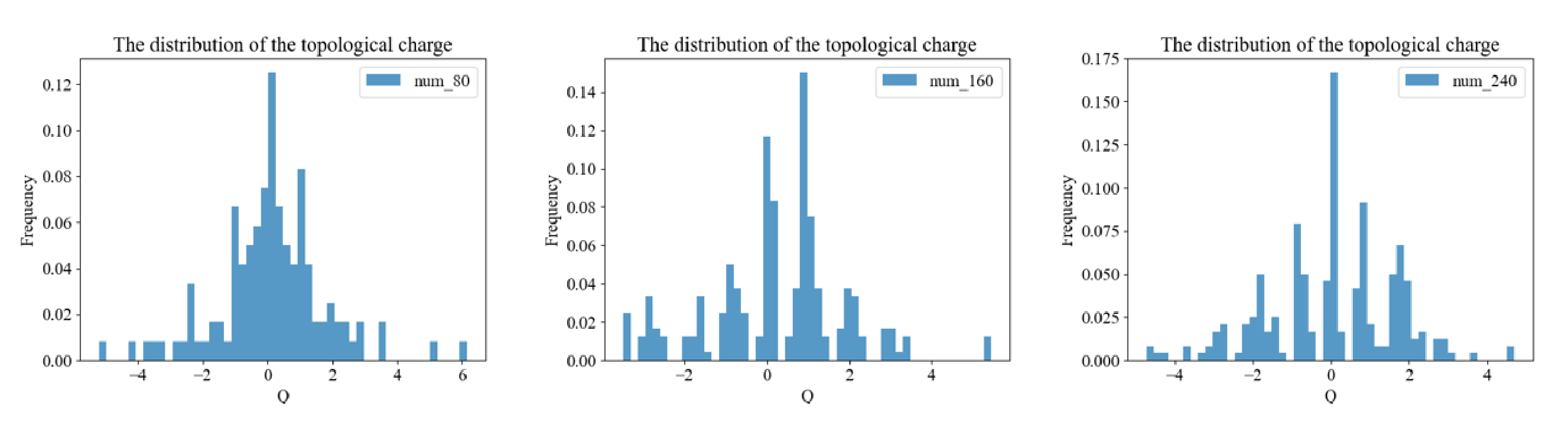}
\caption{\label{figure_4_ref}Comparison of models with 80, 160, 240 training data.}
\end{figure}

For ML, we consider applying 1D M-DCGAN to generate the topological charge used the original MC data as training data. First of all, we need to determine the suitable data volume of training data. We tested the training processes with training data volumes of 80, 160 and 240 by dividing the training data into 2, 2 and 3 groups respectively to train the model. The distributions of topological charge generated by the model trained with different amounts of data are shown in the Fig.~\ref{figure_4_ref}. It can be found that the distribution of the middle subplot is mainly distributed near integers compared with the left subplot, but its peak of the distribution dose not appears at the position of integer zero. Further the distribution of the right subplot is mainly centered on the integers and roughly symmetrical about the integer zero. Therefore, it is obtained that the model trains better as the volume of training data increases. By the experience, we prefer to use 300 MC data divided equally into 3 groups for our train scheme.

\begin{figure}[htb]
\includegraphics[width=0.5\textwidth]{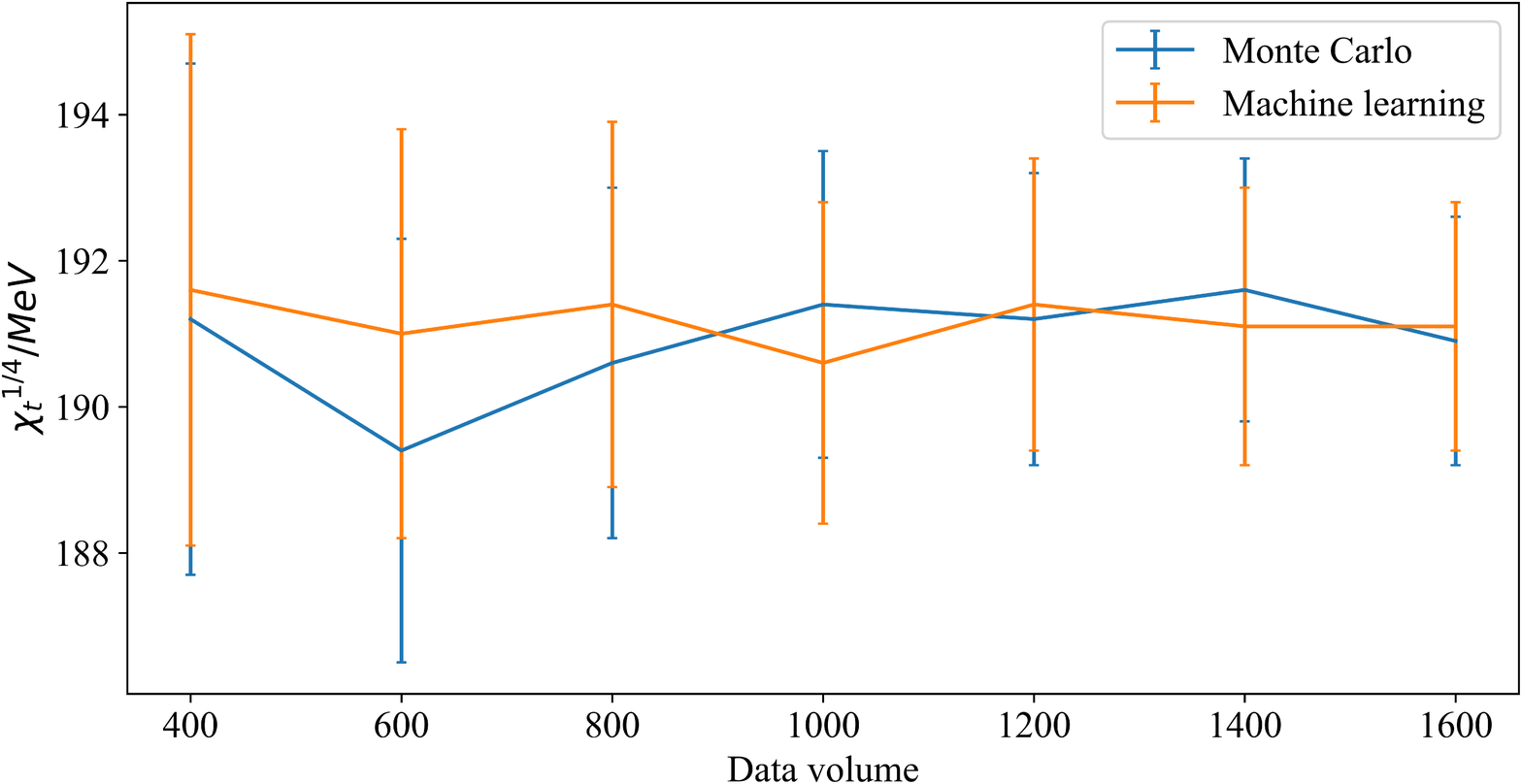}
\caption{\label{figure_5_ref}The fourth root of the topological susceptibility ${\chi_t}^{1/4}$ for MC and ML under different data volume.}
\end{figure}

For our study, the most important thing we noticed is the accuracy of the physics results for MC and ML. It can be observed from Tab.~\ref{MC and ML} and Fig.~\ref{figure_5_ref} that the data error gradually decreases as the volume of data increases for both MC and ML where 300 training data were applied to train 1D M-DCGAN and then 1600 output data were obtained.

\begin{table}[htb]
\caption{\label{MC and ML}The fourth root of the topological susceptibility ${\chi_t}^{1/4}$ for MC and ML.}
\begin{ruledtabular}
\begin{tabular}{ccc}
\textrm{Data volume}&
\textrm{${\chi_t}^{1/4}(MC)/MeV$}&
\textrm{${\chi_t}^{1/4}(ML)/MeV$}\\
\colrule
400 & $191.2\pm3.5$&$191.6\pm3.5$\\
600 & $189.4\pm2.9$&$191.0\pm2.8$\\
800 & $190.6\pm2.4$&$191.4\pm2.5$\\
1000 & $191.4\pm2.1$&$190.6\pm2.2$\\
1200 & $191.2\pm2.0$&$191.4\pm2.0$\\
1400 & $191.6\pm1.8$&$191.1\pm1.9$\\
1600 & $190.9\pm1.7$&$191.1\pm1.7$\\
\end{tabular}
\end{ruledtabular}
\end{table}

When we compare the methods of MC and ML, it is found that the error of the results by the ML is consistent with that by the MC. As shown in Tab.~\ref{time and storage}, the ML model after 300 data training can generate the results as the MC does where both the 1600 data numbers were simulated respectively. And the time cost and storage for the ML also seems reasonable. Therefore, we can apply the ML scheme to generate suitable data based on the MC simulations to deal with the data error and to estimate the results efficiently. 

\begin{figure}[htb]
\includegraphics[width=0.5\textwidth]{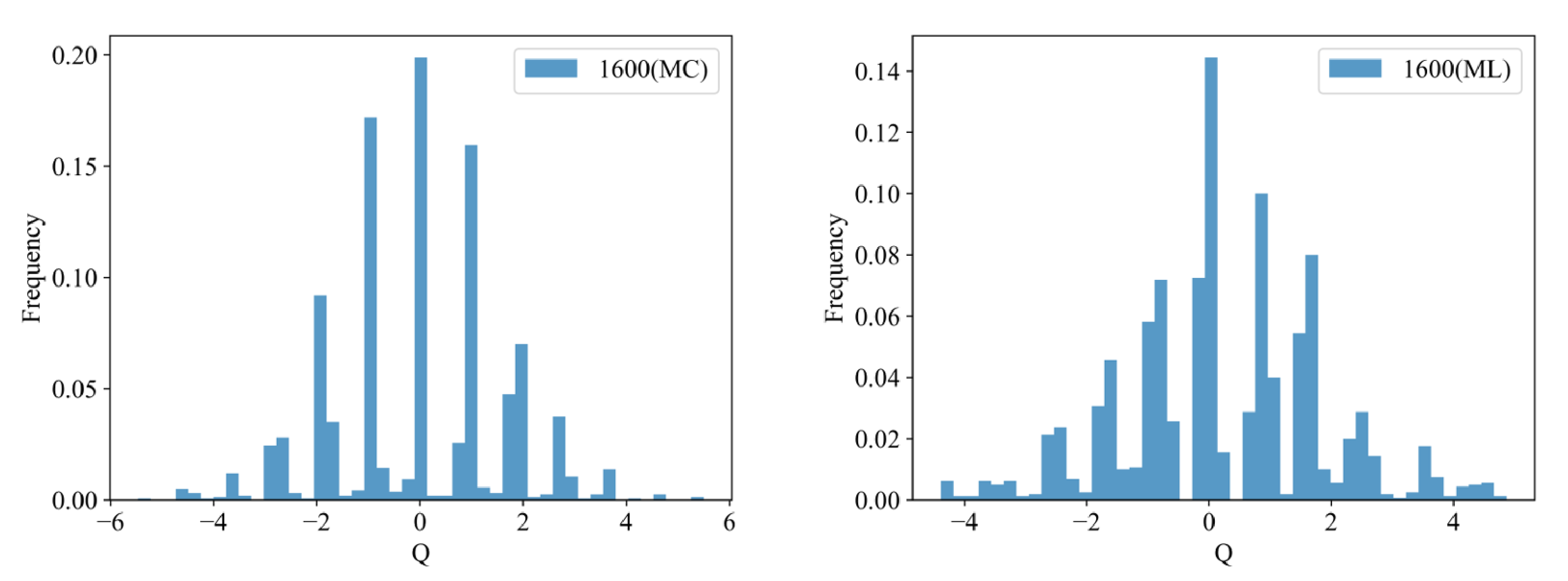}
\caption{\label{figure_6_ref}The distributions of topological charge for MC and ML.}
\end{figure}

Furthermore, the distributions of topological charge for 1600 data generated by the MC and ML methods are shown in Fig.~\ref{figure_6_ref}. It can be observed that both distributions have good symmetry, and their data are discretely distributed at integers. These characteristics are consistent with features of the topological charge. In addition, the integrated autocorrelation time of 1600 data for ML is 0.42, which indicates that these data are independent.

\begin{table}[htb]
\caption{\label{time and storage}A compare of the MC and ML methods for their simulations when 1600 data numbers were used. The time and storage of ML data incorporate the effects of 300 data  training process. }
\begin{ruledtabular}
\begin{tabular}{cccc}
\textrm{Method}&
\textrm{Time/h}&
\textrm{Storage/MB}&
\textrm{${\chi_t}^{1/4}/MeV$}\\
\colrule
MC & 136& 18230& $190.9\pm1.7$ \\
ML & 26& 3429& $191.1\pm1.7$\\
\end{tabular}
\end{ruledtabular}
\end{table}

\subsection{Generation of the topological charge density}
\begin{figure}[htb]
\includegraphics[width=0.5\textwidth]{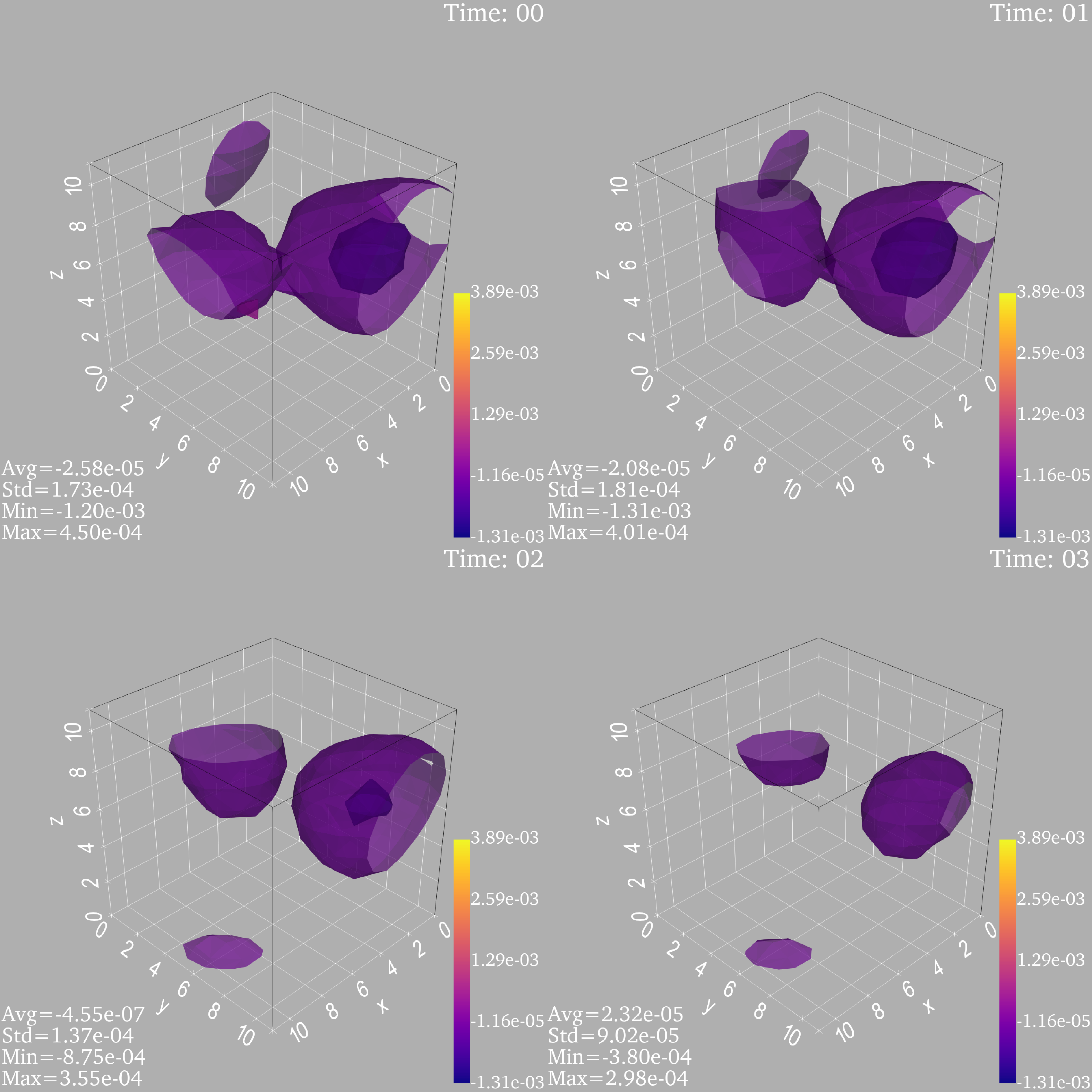}
\caption{\label{figure_7_ref}The images of topological charge density simulated by MC. The four sub-panels correspond to four time slides.}
\end{figure}

Next, the 4D M-DCGAN frame is applied to generate the topological charge density directly in lattice QCD by 300 training data of topological charge density with $N_{flow}=600$ used to train the model of 4D M-DCGAN. The input of generator for 4D M-DCGAN is a random latent variable tensor with shape [-1,100], and the output is a tensor with shape [-1,1,24,12,12,12]. For example, we need to generate a topological charge density tensor with the shape [24,12,12,12]. The input shape is [1,100], and the output shape is [1,1,24,12,12,12]. Finally, the output needs to be reshaped into the tensor with shape [24,12,12,12]. The visualization is referenced from the paper\cite{vege2019nee} and PyVista \cite{sullivan2019pyvista}. The complete topological charge density is four-dimensional, with one-dimensional temporal component and three spatial components. From the previous definition of the topological charge density, it is found that the topological charge density changes continuously in the time direction in the continuous case. Therefore, it will be found that the topological charge density changes approximately continuously in the time direction in lattice QCD when the lattice spacing is small. The images of topological charge density simulated by pseudo heat bath algorithm and Wilson flow are shown in Fig.~\ref{figure_7_ref}.

\begin{figure}[htb]
\includegraphics[width=0.5\textwidth]{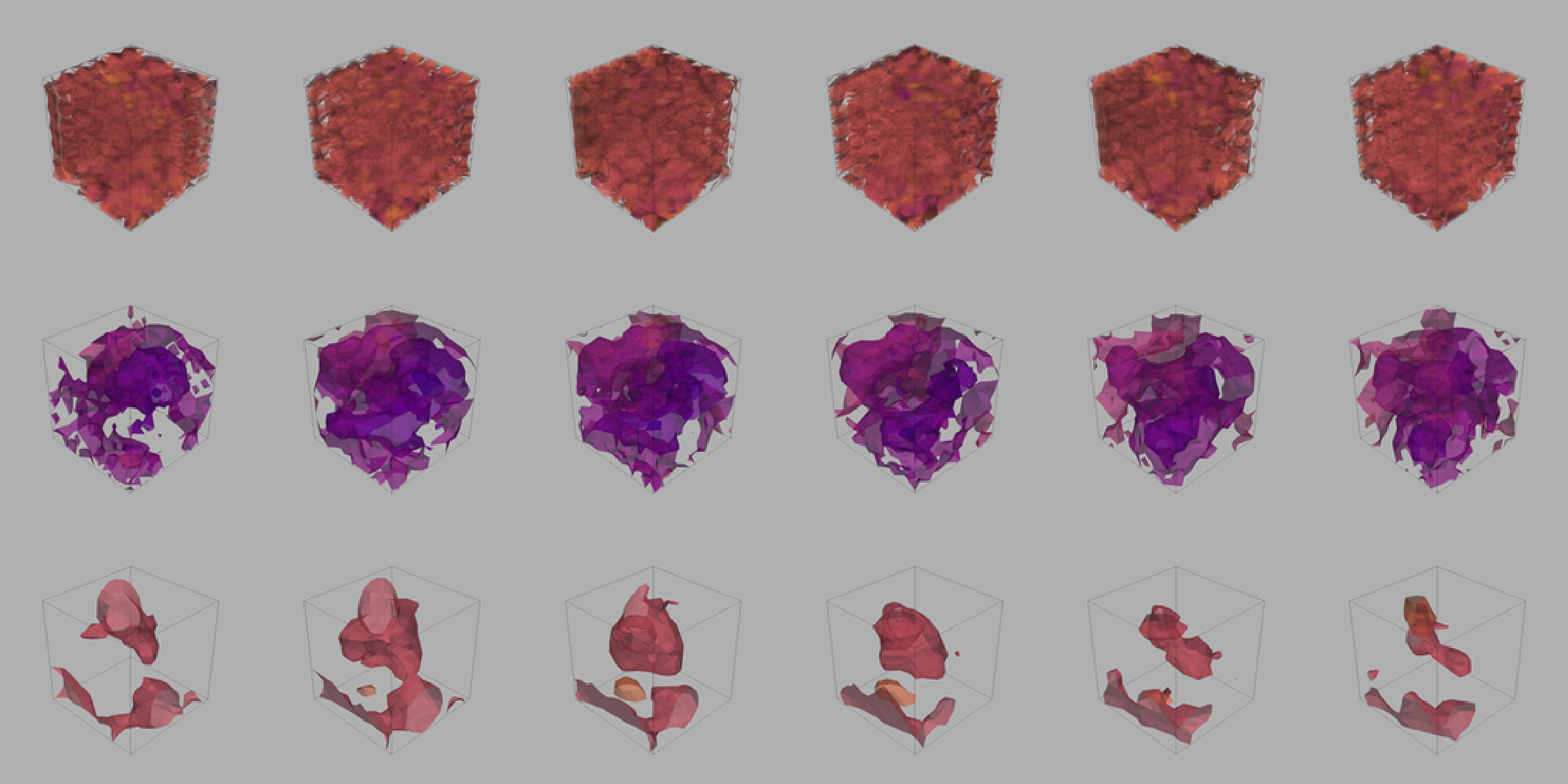}
\caption{\label{figure_8_ref}The sub-images from top to bottom are the topological charge density images generated by the generator when the epoch is 100, 200 and 400, and the sub-images of each row are the pictures of the first 6 time slices of topological charge density. Numerical values are omitted for clarity, and they are discussed later.}
\end{figure}

The images of the topological charge density generated by the generator under different epochs are shown in Fig.~\ref{figure_8_ref}. It is found that 4D M-DCGAN generates clear images from messy images gradually, which means that 4D M-DCGAN does not capture image segments of training data to stitch the images.

Furthermore, the quality of the generated 4D image requires an evaluation process. A painting produced by the generator is handed over to the discriminator for scoring. The discriminator has various evaluation indicators and gets a set of values. These values are then calculated by a certain formula to obtain the final score of the painting. The program is reflected as follows. The generator inputs the generated data to the well-trained discriminator, and the discriminator outputs a vector with various indicators, and then this vector is input into the BCEWithLogitsLoss to generate an evaluation value. It is worth noting that a smaller evaluation value means a higher quality of the generated image. Then we need to convert the evaluation value to a score between zero and one hundred. We introduce the following formula to calculate the score
\begin{equation}
score = 100 \times \frac{{ln{{\bar l}_r} - ln\left( {1 + {l_g}} \right)}}{{ln{{\bar l}_r}}} = 100 \times \frac{{{{ln}}[{{\bar l}_r}/\left( {1 + {l_g}} \right)]}}{{ln{{\bar l}_r}}},
\end{equation}
where ${\bar{l}}_r$ is the average of the evaluation values of 300 random tensors with a size of $Nt\times Nx\times Ny\times Nz$ under the discriminator and BCEWithLogitsLoss, and $l_g$ is the evaluation value of a tensor with the same size as the topological charge density generated by a certain method. The reason for using logarithm in the formula is that ${\bar{l}}_r$ reaches the order of millions. We need to use logarithm to reduce the difference between $l_g$ and ${\bar{l}}_r$, otherwise the real data score will be concentrated between 99 and 100. The $ln\left(1+l_g\right)$ is used to ensure that the score is 100 when $l_g$ is zero. In addition, the criteria for using the score to judge the quality of the data will be given later in conjunction with the discriminator.

\begin{figure}[htb]
\includegraphics[width=0.5\textwidth]{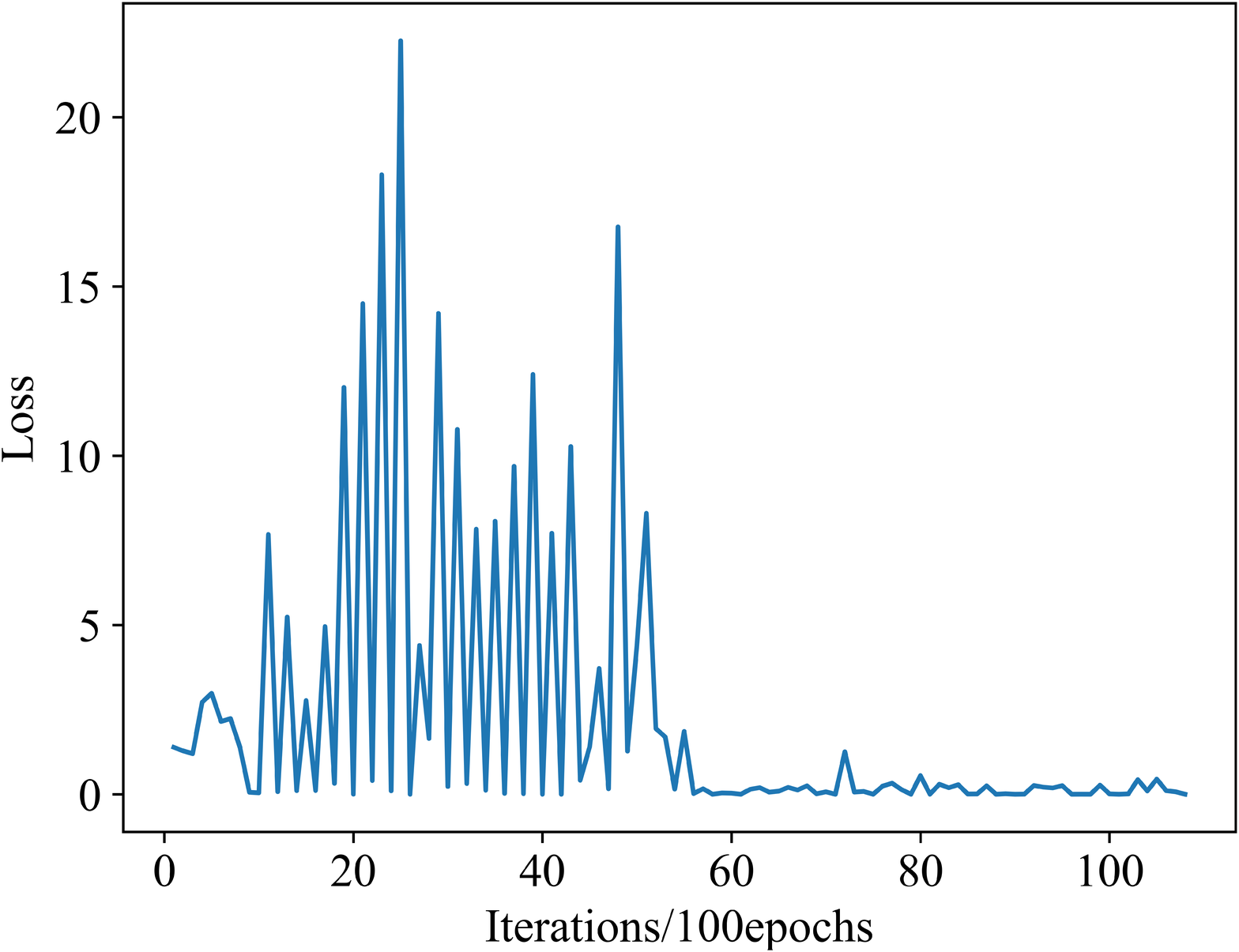}
\caption{\label{figure_9_ref}The losses of discriminator in different training epochs.}
\end{figure}

\begin{figure}[htb]
\includegraphics[width=0.4\textwidth]{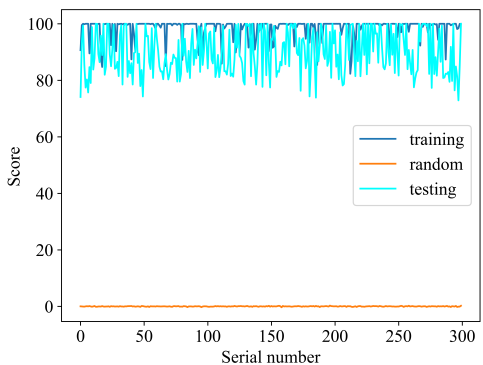}
\caption{\label{figure_10_ref}The scores obtained by training data, test data and random data respectively.}
\end{figure}

As discussed above, the construction method of score requires a well-trained discriminator. Therefore, a well-trained discriminator needs to be picked out. The loss of discriminator is shown in Fig.~\ref{figure_9_ref}. The smaller the loss, the stronger the discrimination ability of the discriminator. It can be found that the loss of the discriminator is not only small but also does not change much after 6000 epochs. Therefore, a well-trained discriminator can be selected after 6000 epochs. Furthermore, we need to determine the score standard of the generated data after we select the discriminator. We selected another 300 topological charge density data calculated by the MC as the testing data. Then we input the training data, testing data and random data into the discriminator, and get the score according to the method mentioned before. The result is shown in Fig.~\ref{figure_10_ref}. It can be observed that the discriminator distinguishes topological charge density data from random data. It can be found from Fig.~\ref{figure_10_ref} that the scores of the real data which include the training data and testing data are in the range of 80 to 100, so the generated data will be credible enough if the score is also in this range.

\begin{figure}[htb]
\includegraphics[width=0.5\textwidth]{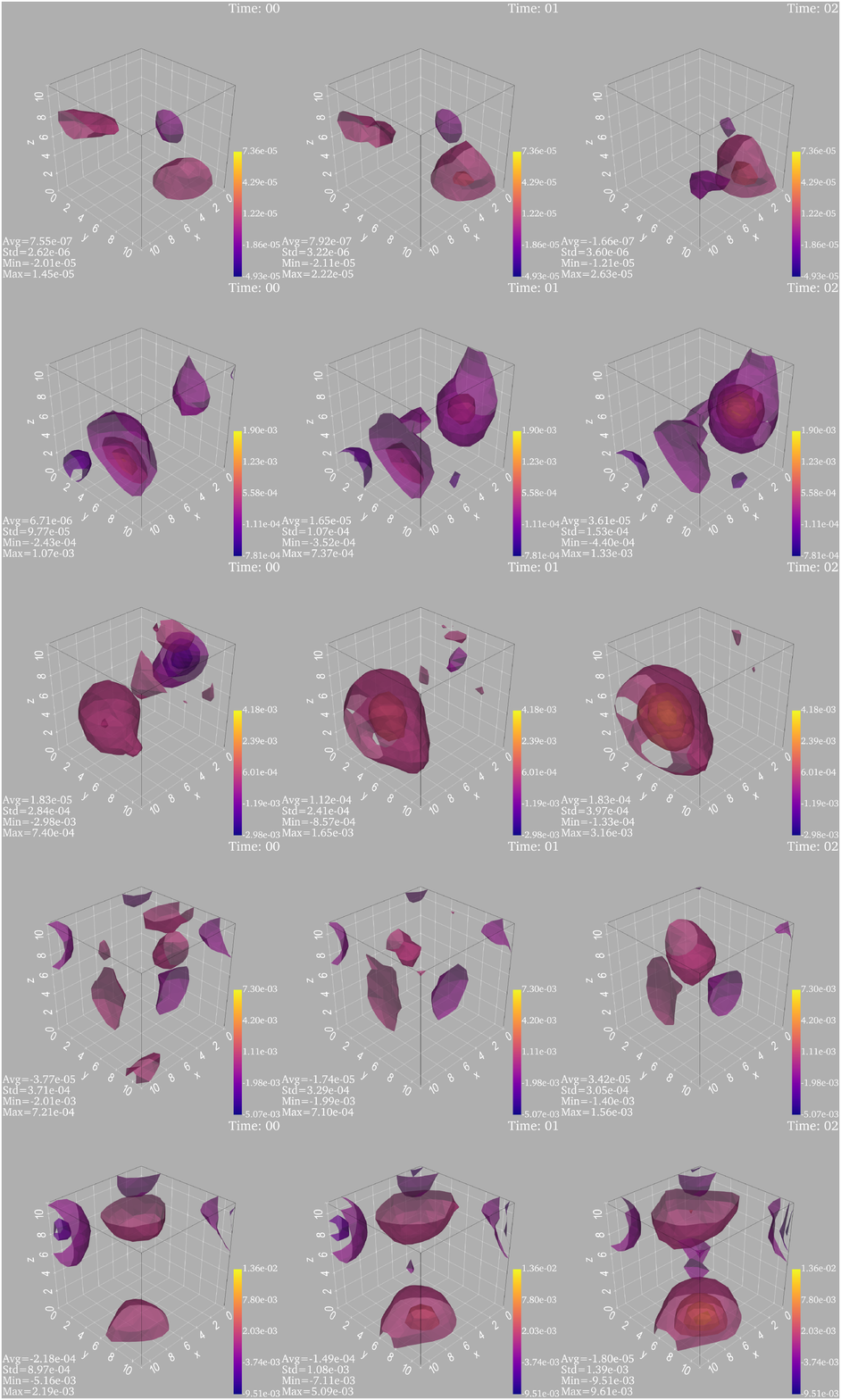}
\caption{\label{figure_11_ref}Images of topological charge density for different values of topological charge. Each row shows the first three sub-images of a topological charge density in time direction. The sub-images from top to bottom correspond to topological charge being 0.02, 1.15, 2.33, 2.98, 3.84 respectively.}
\end{figure}

Some images of topological charge density with different topological charges generated by 4D M-DCGAN are shown in Fig.~\ref{figure_11_ref}. Scores of these topological charge density data are shown in Tab.~\ref{scores}. It can be found that these scores are all larger than 80, which indicates that the generated 4D images are of high quality, or that the generated data of the topological charge density are consistent with the original training data. It is observed from Fig.~\ref{figure_10_ref} and Fig.~\ref{figure_11_ref} that 4D M-DCGAN not only generates images of topological charge density with good shape like real data in Fig.~\ref{figure_7_ref}, but also has a good performance on the continuity of topological charge density in the time direction and the value of topological charge.

\begin{table}[htb]
\caption{\label{scores}The scores of topological charge density with different topological charges under well-trained discriminator and BCEWithLogitsLoss.}
\begin{ruledtabular}
\begin{tabular}{cccccc}
\textrm{topological charge}&0.02&1.15&2.33&2.98&3.84\\
\colrule
\textrm{score}&96&100&95&100&87\\
\end{tabular}
\end{ruledtabular}
\end{table}

\section{Conclusion}
In this paper, the topological quantities of lattice QCD have been studied by a MC simulation mixed with the M-DCGAN frame. Then we have applied an 1D M-DCGAN to generate the distribution of topological charge and a 4D M-DCGAN to calculate the data of topological charge density to show their potentials for applications of ML technique in lattice QCD. By our experience, the conclusions are as follows.

Firstly, compared with the MC with Wilson flow, ML technique with our 1D M-DCGAN scheme shows its efficiency for the MC simulations. The data generated by the 1D M-DCGAN trained with 3 sets of 100 data are more accurate than the corresponding data by the MC simulation as the topological susceptibility is concerned. 

Secondly, it is found from different training epochs that 4D M-DCGAN generates clear images from messy images step by step gradually instead of it simply capturing image segments of training data to stitch the images. Furthermore, the quality of images of topological charge density by the 4D M-DCGAN generator is generally high through the evaluation of well-trained discriminator and BCEWithLogitsLoss. More importantly, the 4D M-DCGAN not only generates images of topological charge density with good shape consistent with that of the image obtained by MC, but also has a good performance on the continuity of topological charge density in the time direction and their values of topological charge centered around integers.

Finally, we hope that the M-DCGAN scheme can be applied to study other physics problems in the lattice QCD to help us for simulating some interesting quantities more efficiently.

\textbf{Acknowledgments.} This work was done based on the Chroma applied to simulate the configuration of the lattice gauge field. We are grateful to the relevant contributors to the Chroma. This article is to be published in Chinese Physics C.

\bibliography{ref}

\end{document}